\documentclass{ws-ijmpe}
 \usepackage{amsmath}
 \usepackage{mathrsfs}
 \usepackage{bm}

\begin{document}

\markboth{J. M. Yao et al.,}{Low-lying states in $^{30}$Mg: a beyond relativistic mean-field investigation}

\catchline{}{}{}{}{}

\title{Low-lying states in $^{30}$Mg: a beyond relativistic mean-field investigation}

\author{\footnotesize J. M. Yao, Z. X. Li, J. Xiang, H. Mei}

\address{School of Physical Science and Technology,\\
 Southwest University, Chongqing 400715 China}
\author{\footnotesize J. Meng\footnote{mengj@pku.edu.cn}}

\address{State Key Laboratory of Nuclear Physics and Technology,\\
  School of Physics, Peking University, Beijing 100871, China\\
 School of Physics and Nuclear Energy Engineering, \\
 Beihang University, Beijing 100191, China}

\maketitle


\begin{abstract}
The recently developed model of three-dimensional angular momentum
projection plus generator coordinate method on top of triaxial
relativistic mean-field states has been applied to study the low-lying states
of $^{30}$Mg. The effects of triaxiality on the low-energy spectra and E0 and E2 transitions
are examined.
\end{abstract}

\section{Introduction}

\begin{figure}[th]
\centerline{\psfig{file=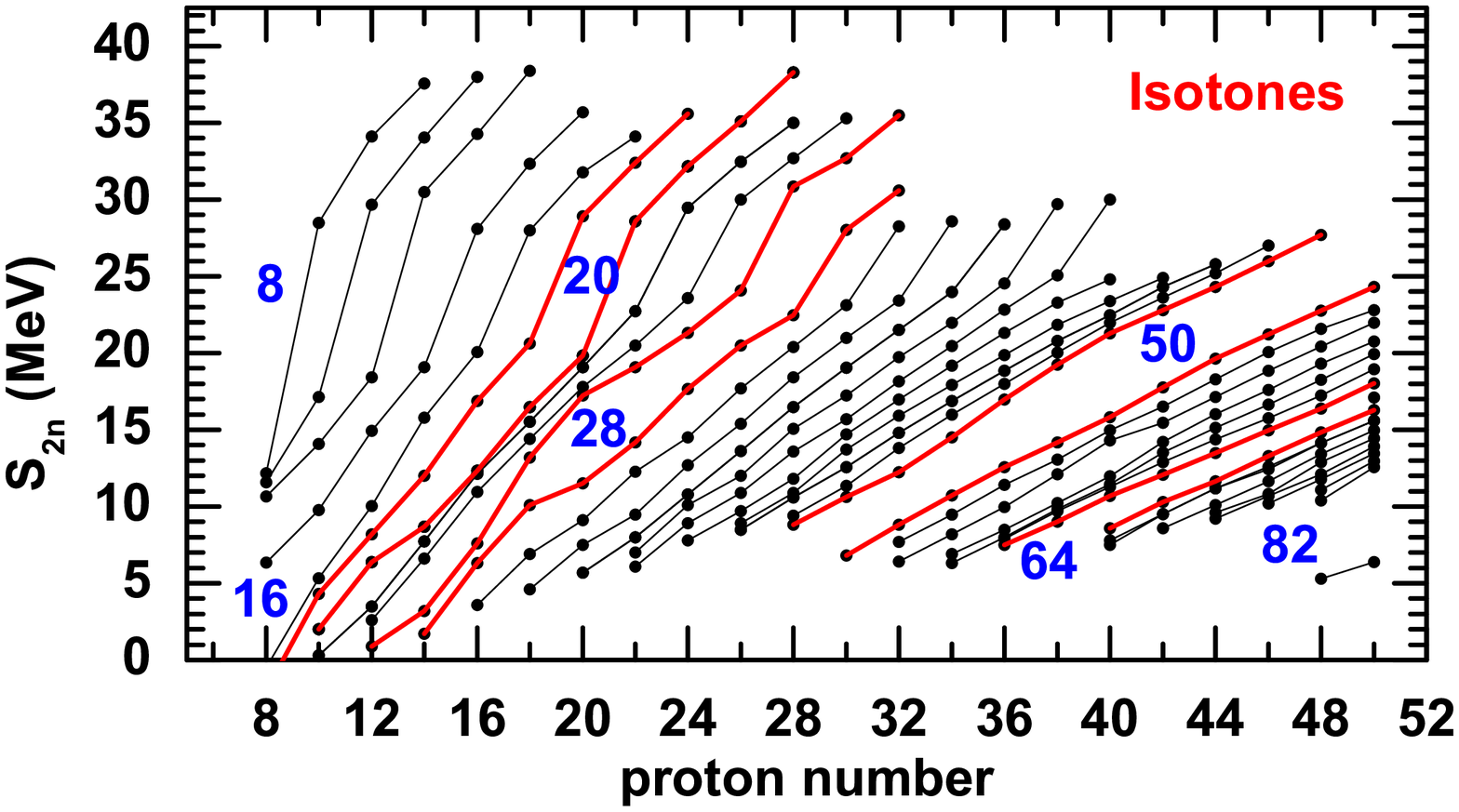,width=7cm}}
\centerline{\psfig{file=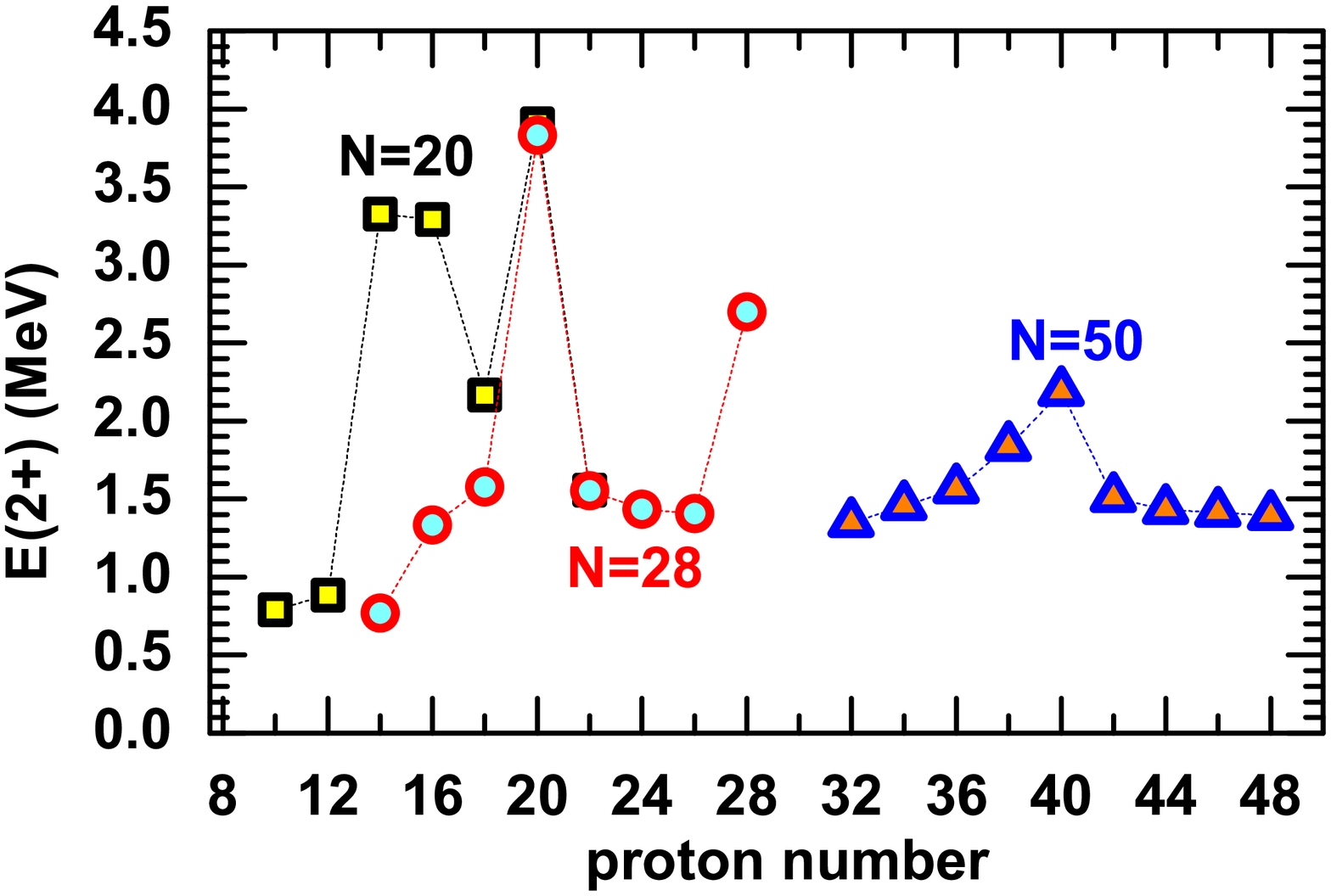,width=7cm} }
 \caption{(Upper panel) Two-neutron
separation energy $S_{2n}$ as a function of proton number for
different isotones. (Lower panel) Systematic variation of the first
excited $2^+$ state $E_x(2^+_1)$ in isotones with $N=20, 28$ and
$50$. All experimental data are taken from Ref.[4].} \label{fig1}
\end{figure}

Radioactive nuclear beam facilities and gamma ray detectors have in
recent years allowed one to study spectroscopy of the low-lying
excited states for exotic nuclei. It provides rich information about
the nuclear structure, including evolution of shell structure and
collectivity, nuclear shape and quantum phase
transition~\cite{Casten06,Cejnar10}, decoupling of neutrons and
protons~\cite{Imai04}. Presently, much interest is focused on the
measurement of the energies
of the first $2^+$, or $4^+$ states and of the reduced transition
probabilities ($B(E2)$-values) from the first $2^+$ to the
ground state ($0^+_1$). These are fundamental
quantities which can be used to disclose the nuclear shapes and
shell structure. A high energy of the first excited $2^+$
state and a corresponding low $B(E2;0^+_1\rightarrow2^+_1)$
electromagnetic transition strength, are characteristic signatures
of shell closures and vice versa. In Figure~\ref{fig1}, we plot the
two-neutron separation energy as a function of proton number for
different isotones. Large gaps are found
between two neighbor isotones with neutron number around $N=8, 20, 28, 50$,
etc. In additional, one notices that the gaps, for instance with
$N=20$ and $28$, are decreasing when one goes to neutron-rich
regions. The evolution information  of shell structure is
reflected in nuclear low-lying states, for example, the first $2^+$
excitation energy as shown in the lower part of Figure~\ref{fig1}.
It is seen clearly that the $E_x(2^+_1)$s in $^{32}$Mg and $^{30}$Ne
are much lower than those in other $N=20$ isotones, indicating the
erosion of magic number $N=20$. Similar situation has been shown in
$N=28$ isotones as well. Therefore, the study of low-lying states provides us an important
way to examine the shell structure in exotic
nuclei and becomes one of the major research fields of the nuclear
structure community.

An alternative approach to the large scale shell model calculations
for nuclear low-lying states is the new beyond mean field
theories~\cite{Bender03}. In recent years several accurate and
efficient models and algorithms have been developed that perform the
restoration of symmetries broken by the static nuclear mean field
and take into account fluctuations around the mean-field minimum.
The most effective approach to configuration mixing calculations is
the generator coordinate method (GCM). With the simplifying
assumption of axial symmetry, GCM configuration mixing of
one-dimensional angular-momentum projected (1DAMP) and even
particle-number projected (PNP) quadrupole-deformed mean-field
states, has become a standard tool in nuclear structure studies with
Skyrme energy density functionals~\cite{Valor00}, the
density-dependent Gogny force~\cite{Guzman02}, and relativistic
density functionals~\cite{Niksic06}. A variety of structure
phenomena has been analyzed using this approach, such as,  the
structure of low-spin deformed and superdeformed collective
states~\cite{Guzman00,Bender03prc,Bender04prc}, shape coexistence in
Kr and Pb isotopes~\cite{Guzman04prc,Bender06prc}, shell closures in
the neutron-rich Ca, Ti and Cr isotopes~\cite{Rodriguez07} and shape
transition in Nd isotopes~\cite{Niksic07PRL,Rodriguez08}.

Recently, the 1DAMP+GCM framework has been extended to include
triaxial shapes, which makes it possible to study the nuclear
low-lying states with the consideration of effects from restoration
of rotation symmetry in full Euler space and shape fluctuation in
$\beta$-$\gamma$ plane. Along this direction, the 3DAMP+GCM models
with PNP have been developed, based on the self-consistent
Hartree-Fock-Bogoliubov approach with Skyrme forces~\cite{Bender08}
and the Gogny force~\cite{Rodriguez10}. In these models, seven
degrees of freedom (three Euler angles + two deformation parameters
+ two gauge angles for protons and neutrons) should be considered,
which makes the calculations very time-consuming. Therefore, only
illustrative calculations have been done with these models. In the
mean time, the 3DAMP+GCM model on top of the relativistic mean-field
has also been developed~\cite{Yao08,Yao09,Yao10}. In the
relativistic 3DAMP+GCM model, the prescription given in
Refs.~\cite{Hara82,Bonche90} has been used to restore approximately
the correct mean values of the nucleon numbers by adding constraint
terms for nucleon numbers to the transition energy functional.

In this report, the recently developed relativistic 3DAMP+GCM model will be
outlined and its applications to the low-lying states of $^{30}$Mg will be presented.
The effects of triaxiality on the low-energy spectra and transitions will be examined.

 \section{The relativistic 3DAMP+GCM model}

 In the relativistic 3DAMP+GCM model, we first perform
 the triaxial relativistic point-coupling model plus BCS calculations
 with quadratic constraints on the mass quadrupole
 moments by minimizing the following energy functional
 \begin{equation}
 \label{eq1}
  E^\prime[\rho_i, j^\mu_i] = E[\rho_i, j^\mu_i] + \sum_{\mu=0,2} \dfrac{C_\mu}{2}(\langle\hat Q_{2\mu}\rangle-q_{2\mu})^2,
 \end{equation}
 that is a functional of four types (S, V, TS, TV) of densities and currents. $\langle\hat Q_{2\mu}\rangle$ denotes the expectation
 value of the mass quadrupole operator, $q_{2\mu}$ is the triaxial deformation parameter, and $C_\mu$ is the stiffness constant.
 This procedure generates a large set of highly correlated intrinsic deformed states $\vert
  \Phi(q)\rangle$. There are successful non-linear versions, PC-F1\cite{Burvenich02}, PC-PK1\cite{Zhao10} and density-dependent version, DD-PC1\cite{Niksic08}
  of relativistic energy functionals that can be used in Eq.(\ref{eq1}).
  The pairing correlations, for open-shell nuclei,
  are taken into account by augmenting the following pairing energy
  functional,
  \begin{equation}
  E[\kappa]=-\sum_{\tau}\int d\mathbf{r} \dfrac{V_\tau}{4} \kappa^\ast_\tau(\mathbf{r} )\kappa_\tau(\mathbf{r})
   \end{equation}
  with separately adjustable pairing strengths $V_{p/n}$ for protons and neutrons. The anomalous density (pairing tensor) $\kappa_\tau(\mathbf{r})$
  is determined by the occupation probability coefficient $v_k$ and density $\rho_k(\mathbf{r})$ of single-particle state,
  \begin{equation}
  \kappa(\mathbf{r})
  =-2\sum_{k>0}f_ku_kv_k\rho_k(\mathbf{r}),
  \end{equation}
 where $f_k$ is an energy-dependent smooth cutoff factor used to simulate the effects of finite-range.
 Alternatively, one can use a separable pairing force~\cite{Tian09} in the pairing channel. The results of
 low-lying states with the separable pairing force will be given elsewhere.

The nuclear wave function $ \vert\Psi^{JM}_\alpha\rangle$ with good angular momentum and shape fluctuation is obtained by projecting the
intrinsic states $\vert\Phi(\beta,\gamma)\rangle$ onto good angular momentum (K-mixing) and
performing GCM calculations (configuration mixing),
   \begin{equation}
     \label{TrialWF}
     \vert\Psi^{JM}_\alpha\rangle
     =\int d\beta d\gamma
      \sum_{K\geq0}f^{JK}_{\alpha}(\beta,\gamma) \frac{1}{(1+\delta_{K0})}
      [\hat P^J_{MK}+(-1)^J\hat P^J_{M-K}]\vert\Phi(\beta,\gamma)\rangle
    \end{equation}
The weight functions $f^{JK}_{\alpha}$ are determined from the
solution of Hill-Wheeler-Griffin (HWG) integral equation
 \begin{equation}
 \label{HWEq}
 \int dq^\prime\sum_{K^\prime\geq0}
 \left[\mathscr{H}^J_{KK^\prime}(q,q^\prime)
 - E^J_\alpha\mathscr{N}^J_{KK^\prime}(q,q^\prime)\right]
  f^{JK^\prime}_\alpha(q^\prime)=0,
 \end{equation}
 where $\mathscr{H}$ and $\mathscr{N}$ are the angular-momentum projected GCM
 kernel matrices of the Hamiltonian and the Norm, respectively.
 $q$ is the abbreviation of triaxial deformation parameters $(\beta, \gamma)$.


 The electromagnetic moments and transition strengths of low-lying states are
 evaluated with the nuclear wave function $\vert\Psi^{JM}_\alpha\rangle$.
 More details about the relativistic 3DAMP+GCM model can be found in Refs.~\cite{Yao09,Yao10}.

%
%
%
%
%

\section{Low-lying states in $^{30}$Mg}

\begin{figure}[th]
\centerline{\psfig{file=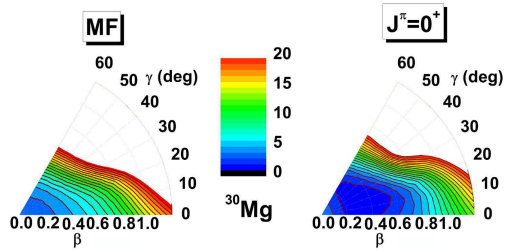,width=7cm}} \vspace*{8pt}
\vspace{-1cm}
\caption{Self-consistent RMF+BCS triaxial quadrupole energy surface
and angular-momentum-projected energy surface with $J^\pi=0^+$ from the
3DAMP+GCM calculations with the effective interaction PC-F1.
All energies are normalized with respect to the binding energy of the absolute minimum.
The contours join points on the surface with the same energy.
The difference between neighboring contours is 1.0 MeV.}
\label{fig2}
\end{figure}

\begin{figure}[th]
\centerline{\psfig{file=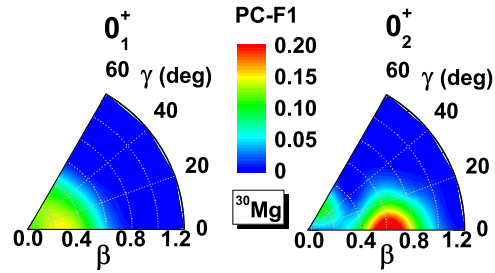,width=7cm}} \vspace*{8pt}
\caption{Contour plots of the probability distribution of collective wave functions
for the lowest two $0^+$ states.}
\label{fig3}
\end{figure}

Figure~\ref{fig2} shows the self-consistent RMF+BCS triaxial quadrupole energy surface
and angular-momentum-projected energy surface  with $J^\pi=0^+$ from the
3DAMP+GCM calculations with the effective interaction PC-F1. It is seen that
there is an evident near-spherical minimum in mean-field energy surface of $^{30}$Mg.
However, this minimum shifts to a relative large deformed shape with $\beta=0.4$ and $\gamma=20^\circ$.
The dynamic correlation energy from the restoration of rotational symmetry is about 2.7 MeV.
The projected energy surface becomes soft in both $\beta$ and $\gamma$ directions around the minimum, indicating
the existence of large shape fluctuations. This phenomenon can be seen from the probability distribution of
ground state in Fig.~\ref{fig3}.

To examine the effects of triaxiality on the low-energy spectra and transitions, we present the
spectroscopic results from both 1D- (frozen $\gamma=0^\circ, 180^\circ$) and 3D-AMP+GCM calculations
in Table 1, where the results of 1DAMP+GCM calculations with the non-relativistic Gogny force and
corresponding experimental data~\cite{Schwerdtfeger09} are given for comparison. It shows that the inclusion
of triaxiality in the 3DAMP+GCM calculations with the PC-F1 brings the results of $E_x(2^+_1), E_x(0^+_2)$
and $E0$ transition strength $\rho^2_{21}(E0)\times10^3$ more closer to the data.

\begin{table}[h!]
\label{tab1} \tbl{Results from both 1D and 3DAMP+GCM calculations
with the PC-F1 force. Both the non-relativistic calculation results
with Gogny force and experimental data are taken from Ref.[28].}
{\begin{tabular}{cccccc} \toprule
         & $E_x(2^+_1)$ & $E_x(0^+_2)$ & $\rho^2_{21}(E0)\times10^3$ & $B(E2;0^+_1\rightarrow2^+_1)$& $B(E2;0^+_2\rightarrow2^+_1)$\\
         & (MeV) & (MeV) &   &  (e$^2$fm$^4$) & (e$^2$fm$^4$)    \\
 \colrule
Exp.       & 1.482        & 1.789        &  26.2$\pm$7.5 & 241(31) & 53(6) \\
3D(PC-F1)  & 1.721        & 2.864        &   24.72       & 277     &  68  \\
1D(PC-F1)  & 1.882        & 3.275        &  15.56        & 257     &  47   \\
1D(Gogny-D1S)& 2.03       & 2.11         &  46           & 334.6   & 181.5   \\
\botrule
\end{tabular}}
\end{table}

 \section{Summary}

In conclusion, the recently developed relativistic 3DAMP+GCM model has been
outlined and its applications to the low-lying states of $^{30}$Mg has been presented.
It has been found that $^{30}$Mg has large shape fluctuations in both $\beta$ and $\gamma$
directions around the triaxially deformed minimum.
The effects of triaxiality on the low-energy spectra and transitions have been shown
to be important to reproduce the corresponding experimental data.

\section*{Acknowledgements}

J.M.Y would like to express his deep gratitude to all his
collaborators, in particular to K. Hagino, Z. P. Li, D. Pena
Arteaga, P. Ring, and D. Vretenar for the active collaboration over
many years on the topics addressed in this paper.
This work has been supported in part by the Major State 973
Program 2007CB815000 and the National Natural Science Foundation of
China under Grant Nos. 10947013, 10975008, 10705004 and 10775004, the Southwest University
Initial Research Foundation Grant to Doctor (No. SWU109011).


\end{document}